\documentclass[11pt]{amsart}
\usepackage{geometry}                
\usepackage{graphicx}
\usepackage{amssymb}
\usepackage{epstopdf}
\newcommand{\curl}{\mbox{curl}}
\newcommand{\R}{\mathbb{R}}
\newcommand{\Z}{\mathbb{Z}}
\newcommand{\T}{\mathbb{T}}
\renewcommand{\S}{\mathbb{S}}

\title[Invariant tori for the pressure jump Hamiltonian]{Invariant tori  for the pressure-jump Hamiltonian}
\author{R.S.MacKay\\ \\
in memory of Bob Dewar}
\address{Mathematics Institute, University of Warwick, Coventry CV4 7AL, UK}
\email{R.S.MacKay@warwick.ac.uk}
\date{\today}                                           

\begin{document}
\begin{abstract}
A major achievement of Dewar and coworkers is the SPEC code to construct stepped-pressure equilibria in magnetohydrostatics without axisymmetry.  
Their existence had been proved by Bruno and Laurence.
As part of the procedure of Bruno and Laurence, it is required to solve the Hamilton-Jacobi equation for a magnetic potential on the outside of an interface given the field on the inside and the pressure-jump across the interface.  For non-axisymmetric interface, it was understood that solutions with insufficiently irrational rotational transform might not exist, and examples have been given for which there are no solutions at all for large enough pressure-jump.  The present paper gives a method to compute regions in the phase space for the pressure-jump Hamiltonian through which no invariant tori pass.  The paper also shows how to present the results as regions in the space of pressure-jumps and outer rotational transform for which there is no solution of the Hamilton-Jacobi equation.  The method is expected to reach arbitrarily close to the full non-existence region with enough computational work, so what is left over can be relied on to be mostly invariant tori.  The paper also brings to attention a class of metrics on tori that are not necessarily axisymmetric yet have integrable geodesic flow. They could give interfaces with solutions for all but finitely many rotational transforms.
\end{abstract}

\maketitle

\section{Introduction}
Bob Dewar made many contributions to plasma physics and Hamiltonian dynamics.  I got to know him when I moved into Theory wing as a PhD student in Princeton Plasma Physics Laboratory.  Deeper research interactions with him, however, date from 2004 when he invited me to join the Complex Open Systems network (COSnet).  He was very excited by the idea that one could extend the mathematical construction by \cite{BL} of stepped pressure magnetohydrostatic (MHS) equilibria (which built on \cite{B+}) to a computational method that could produce useable solutions.  Indeed, he achieved this in \cite{H+} and the SPEC code.  The solutions consist of Beltrami zones (magnetic field $B$ satisfying $\curl\, B = \mu B$ for some constant $\mu$ in the zone) separated by toroidal interfaces that are invariant under both the limiting inside and outside magnetic fields, supporting a current sheet and a pressure jump.

When he explained this to me during a visit to him in Canberra in 2005, my instant reaction was that pressure jumps and the associated current sheets would be smoothed out by diffusive effects, so it would be much nicer to find MHS fields with continuous pressure and bounded current density, but still consisting mainly of Beltrami zones.  If the pressure is not constant, an obvious requirement is infinitely many Beltrami zones and uncountably many interfaces, thus perhaps hard to implement numerically, but still mathematically and physically interesting.  Thus began a discussion that we kept going until his end.  We didn't reach a conclusion, but this sounded a plausible way to realise Grad's claim that there are MHS equilibria with infinitely many zones of constant pressure and continuous pressure \cite{G}.

For this paper, I return to the case of constructing MHS solutions with pressure jumps.   A key step in \cite{BL} was to solve the Hamilton-Jacobi (HJ) equation for a magnetic potential, giving the magnetic field on the outside of a toroidal interface, in response to (i) given field on the inside, (ii) specified pressure jump, and (iii) one further parameter that one can take to be the rotational transform of the outer field.  This was revisited by \cite{M+}.  Although SPEC takes a different approach to constructing stepped-pressure equilibria \cite{H+}, an understanding of which interfaces can support how much pressure jump is an important ingredient that still remains to be automated.
In numerical implementations of SPEC with outer boundary and interfaces not too far from axisymmetric, a satisfactory stepped-pressure equilibrium was usually found \cite{H+}, but if they were too far from being axisymmetric, the method would not converge \cite{Q+}.  

The present paper has two purposes.  The first is to give a method to compute regions in the space of pressure-jumps and rotational transforms for which there is no solution of the HJ equation.  The method is an application of \cite{M} to this problem, extended by the idea of \cite{J}.  
To elaborate, \cite{KS} already gave examples of interface for which they proved there is no solution for any external rotational transform for large enough pressure jump.  Their proof can be seen as a special case of the condition of \cite{M}.
It is plausible that the method is exhaustive, in the sense that as the computation time is taken to infinity the set of parameter values for which existence is not eliminated converges down to the set for which there exists an invariant torus (compare \cite{S,J} where exhaustivity is proved for the case of area-preserving twist maps).

The second purpose  is to highlight a class of integrable metrics on tori that is broader than the axisymmetric ones and so could give non-axisymmetric interfaces for which the Hamilton-Jacobi equation has solutions for all but two rotational transforms (at least for some value of pressure jump).

\section{Pressure-jump Hamiltonian}
We treat MHS fields in a general oriented 3D Riemannian manifold $M$, because it will not cost us much to generalise from ordinary 3D Euclidean space.  The main place the metric appears is in its restriction to an interface.  

A magnetic field is a divergence-free vector field $B$ with respect to the Riemannian volume $\Omega$ for the chosen orientation, i.e.~$i_B\Omega$ is closed (if $M$ has nontrivial cohomology one should add that the flux form $i_B\Omega$ be exact).
Denote the (Riemannian) norm of a vector $B$ by $|B|$; in (contravariant) components, $|B| = \sqrt{g_{ij}B^iB^j}$, with $g $ the Riemannian metric tensor.

Denote the limiting magnetic fields on each side of a smooth interface $\Sigma$ by $B^\pm$.  The fields $B^\pm$ are assumed to be tangent to $\Sigma$.  
The jump condition is
\begin{equation}
\tfrac12|B^+|^2 = \tfrac12|B^-|^2 + [P],
\label{eq:jump}
\end{equation}
where $[P] = P^- -P^+$, the change in pressure (in the opposite direction) across the interface.
By the MHS condition, the current is also tangent to $\Sigma$, so on each side there is a  potential $f^\pm: \Sigma \to \R$ such that
\begin{equation}
(B^\pm)^\flat = df^\pm,
\label{eq:pot}
\end{equation}
where for a vector field $B$, $B^\flat$ is the 1-form $B\cdot dx$ (known as the covariant representation of $B$).   Equivalently, $B^\pm = \nabla f^\pm$, where $\nabla$ is defined using the cotangent metric ($B^{i} = g^{ij} \partial_j f$). The potentials must in general be allowed to be multivalued; formally, one ought to say there are closed 1-forms $\alpha^\pm$ such that $(B^\pm)^\flat = \alpha^\pm$. 

If one supposes $B^-$ to be known and $[P]$ to be specified, the problem of determining $B^+$ reduces to solving
\begin{equation}
\tfrac12 g^{ij}\partial_if^+\partial_jf^+ = \tfrac12 |B^-|^2+[P]
\label{eq:HJ}
\end{equation}
for $f^+$.
This can be recognised as the HJ equation for an invariant Lagrangian graph $p = df(q)$ of the Hamiltonian
\begin{equation}
H(q,p) = \tfrac12 g^{ij}(q)p_ip_j - V(q)
\label{eq:PJHam}
\end{equation}
on $T^*\Sigma$ with ``energy'' $H=[P]$, where $V = \tfrac12 |B^-|^2$ and $p = df^+$.
Here, $q$ denotes a point on $\Sigma$ and $p$ a cotangent vector to $\Sigma$ at $q$.  Note that the sign of $V$ is opposite to that used for many Hamiltonian systems, but it is convenient to keep it that way because $V>0$ (indeed, celestial mechanics tend to use this sign convention for the same reason).

The relevant case is $\Sigma$ diffeomorphic to a 2-torus.  Indeed, if $B^-$ is nowhere zero on $\Sigma$ and $\Sigma$ is bounded,  boundaryless and two-sided, then $\Sigma$ must be a 2-torus.  Thus the invariant Lagrangian graphs in question for $H$ are also 2-tori.  We will refer to them as invariant tori and restrict attention to those that are Lagrangian graphs over $\Sigma$.  The reader is asked to remember this restricted use of the term ``invariant torus''.

There may be many invariant tori.  Indeed if $H$ is smooth enough and has one invariant torus with a smooth enough conjugacy to a Diophantine rotation and non-zero shear, then there are uncountably many such for each nearby value of $[P]$ (there is a Birkhoff normal form in a neighbourhood of such a torus and then KAM theory applies close enough to it, e.g.~\cite{SZ}).  In particular, the shear for invariant tori of (\ref{eq:PJHam}) is always positive so if $B^-$ has smooth enough conjugacy to a Diophantine rotation then we have a suitable invariant torus with $[P]=0$ and hence deduce existence of invariant tori for all sufficiently irrational winding ratios near $\iota^-$ for small enough $[P]$.

On the other hand, for some values of $[P]$ there might be none.  A simple argument rules out all invariant tori with $[P] < -V_{\min}$, where $V_{\min} = \min_{q \in \Sigma} V(q)$, namely, that $|B^+|^2$ would then have to be negative at a minimum of $V$.  In practice, this might not be relevant, as the \cite{BL} construction of stepped pressure solutions works from the inside out and $V > 0$, and one wants $[P]>0$ for each interface, corresponding to a decrease in pressure; but it could be relevant in other contexts.
So henceforth, attention will be restricted to $[P] \ge - V_{\min}$ (indeed, to $[P]> - V_{\min}$, else $B^+=0$ at the minimum).

Of more interest, one might ask what is the set of invariant tori of the pressure-jump Hamiltonian for an interface that is a boundary torus for the field on the inside, i.e.~an invariant torus such that there is an inside-neighbourhood containing no others \cite{MS}.  There might be an interesting renormalisation picture here.

One way to impose that there is at most one solution of the HJ equation is to fix the ``periods'' of the potential $f^-$.  These are the numbers $c_i$ by which $f^+$ changes for one positive circuit in the positive toroidal ($i=t$), respectively poloidal ($i=p$) direction.  By Stokes theorem, $c_p$ represents the current crossing any poloidal section bounded by a poloidal loop in $\Sigma$, including the current in the sheet on $\Sigma$.  Similarly, if $\Sigma$ encloses a closed field line $\gamma$ and we let $c_\gamma  = \int_\gamma B\cdot dx$ then $c_t-c_\gamma$ represents the current crossing any section spanning from $\gamma$ to a toroidal loop on $\Sigma$, again including the current sheet.   Uniqueness of a torus with given periods holds because the difference $\delta f$ of two solutions of (\ref{eq:HJ}), even allowing different $[P]$, satisfies $|\nabla\, \delta f|^2 = $ cst (with the cotangent metric), so with fixed periods, $\delta f = 0$.  Rather than specifying the periods of $f^+$ it is equivalent to specify $[P]$ and the rotational transform $\iota^+$ of $B^+$ (the limit of ratio of number of poloidal revolutions to toroidal ones along any field line of $B^+$ as its length goes to infinity).

In the work of Dewar and coworkers, e.g.~\cite{Q+}, it is usual to take $\iota^+ = \iota^-$ and to be a good irrational.  Choosing a good irrational is essential in view of generic non-existence of invariant tori with rotational transform too close to rationals \cite{M+}, but choosing the inner and outer rotational transforms to be the same, although a sensible starting point, especially for small $[P]$ because of the above application of KAM theory, seems an arbitrary restriction (acknowledged by \cite{H+}) that we shall not require.

\section{Conditions for non-existence of invariant tori}

The pressure-jump problem can be reformulated as a Lagrangian system on $T\Sigma$ with Lagrangian
\begin{equation}
L(q,v) = \tfrac12 |v|^2 +V(q).
\label{eq:Lag}
\end{equation}
It is positive-definite in the velocity $v$, which is related to $p$ by $v^i = g^{ij}p_j$ (so $p = v^\flat$).
A theorem of Weierstrass (see \cite{M}) states that every trajectory on an invariant Lagrangian graph for positive-definite Lagrangian has minimal action with respect to all variations of compact support.  The action of a segment of a trajectory is
\begin{equation}
S = \int_{t_0}^{t_1} L(q(t),\dot{q}(t))\ dt.
\label{eq:action}
\end{equation}
This gives some intuition about regions of phase space through which no invariant tori can pass and can lead to proofs of non-existence regions.  
In particular, it may be cheaper to go round a maximum of $V$ than over it, at least if $[P]$ is not too large (so that the increase in the integral of the kinetic energy from moving the path is smaller than the decrease in $\int V\ dt$).

Another reformulation for given $[P] > -V_{\min}$, is as the geodesics of the {\em Maupertuis metric}
\begin{equation}
ds^2 = \tfrac12([P] + V(q))|dq|^2,
\label{eq:Mau}
\end{equation}
where the time-parametrisation is given subsequently by 
\begin{equation}
dt = \frac{|dq|}{\sqrt{2([P] + V)}} = \frac{ds}{[P] + V}.
\label{eq:time}
\end{equation}
This has the advantage that one can work with the lengths $\ell$ of unparametrised paths $\gamma$:
\begin{equation}
\ell(\gamma) = \int ds = \int \sqrt{\tfrac12([P] + V(q))}\, |dq|.
\label{eq:length}
\end{equation}
If the length of a path between two points can be shortened then the original path was not on an invariant Lagrangian graph.
There are various papers about Riemannian metrics on 2-tori with no invariant tori for their geodesic flow, e.g.~\cite{Ba}. 
In particular, if the metric has a ``big bump'' then it is shorter to go round it than across it, hence there are no invariant tori.  This was also the method used by \cite{KS} to make examples of interfaces that could not support any pressure jump (though note that they took $B^-=0$).
We expand on this idea below.
See also \cite{CPC} for a wide survey of the Riemannian view of Hamiltonian dynamics.

The above results can be turned into a rapid computational method to find regions of non-existence of invariant tori, that Percival and I christened ``Converse KAM theory'' \cite{MP}.  It has been demonstrated on the two-wave Hamiltonian \cite{M}.  An extension from the positive-definite case \cite{M18} has been demonstrated on the three-body problem \cite{KMS} and on magnetic fieldline flow \cite{KMM,MM}.
It would be good to demonstrate it on a typical pressure-jump Hamiltonian and then use it routinely in construction of stepped pressure equilibria.  Here I indicate how this can be done.

There are various ways to formulate Converse KAM theory, depending also on the chosen formulation of the dynamics (Hamiltonian, Lagrangian, geodesic).
One that holds for all three contexts is that if the trajectory of a point has conjugate points then it is not on an invariant torus.  Recall that two points on a trajectory are called {\em conjugate} if there is a non-zero trajectory of the linearised equations with $\delta q=0$ at both ends.  

Specialise to the geodesic context.  If there is $q_0 \in \Sigma$ and initial velocity $v$ in the unit tangent space at $q_0$ such that the geodesic in that direction has conjugate points, then there is no invariant torus containing that geodesic.  This would happen if there is $q_1$ on the geodesic conjugate to $q_0$, but sometimes it can be more efficient to look for a pair of conjugate points $(q_{-1}, q_1)$ on the geodesic through $q_0$ with $q_{\pm 1}$ in opposite directions along it.

For geodesic flows in two dimensions, the linearised equations can be reduced modulo the tangent to the geodesic to looking at just the perpendicular component $w$. It evolves by the Jacobi equation $$w''=-\kappa w,$$ 
where $\kappa$ is the Gaussian curvature for the metric and $'$ denotes derivative with respect to arclength.  Long enough segments of a geodesic wtih positive curvature produce conjugate points.  For example, if $\kappa \ge \pi^2/L^2$ on an arc of length $L$ of a geodesic, then the arc contains conjugate points.
Extending this, large enough regions of positive curvature produce conjugate points for all geodesics through their centre.  As already mentioned, such regions are called ``big bumps''.
In particular, if $\kappa \ge \pi^2/L^2$ for all points within geodesic distance $L/2$ of some point $q_0$ then there are no invariant tori at all.

The Maupertuis metric is conformally equivalent to the original metric on $\Sigma$, so one can compute its curvature using that the curvature for metric $h=e^{2w}g$ is
\begin{equation}
\kappa_h = e^{-2w}(\kappa_g-\Delta_g w), 
\end{equation}
where $\Delta_g$ is the Laplacian for $g$.
So $\kappa_h$ is more positive where $\kappa_g$ is positive and $\Delta_g w$ is negative.  The curvature $\kappa_g$ can be computed as the product of the principal curvatures of $\Sigma$ as a surface in 3-space, so $\kappa_g>0$ corresponds to the ``outboard'' side of $\Sigma$.  The condition $\Delta_g w < 0$ corresponds roughly to being near the maximum of $V$, because
in our case $e^{2w} = \tfrac12([P] + V)$, so $w = \tfrac12 \log \tfrac12([P] + V)$.  Computing $\Delta_g w$ is a bit of a mess, but one can obtain an idea of its sign and size by using $\int_{D_r} \Delta_g w\, dA_g = \int_{\partial D_r} \nabla_g w \cdot dn_g$, where $D_r$ is the disk of radius $r$ (in the metric $g$).  Because $V = \frac12 |B^-|^2$ can be expected to be smallest on the outboard side, however, these two considerations are in opposition (this is expected, because for $[P]=0$ we know there is a solution), so more detailed computation is required to find a large enough region with $\kappa_h > 0$.  On the other hand, 
for $[P]$ large positive, $w$ is approximately constant, so only $\kappa_g$ is relevant.  If the interface contains a disk of some radius $r$ (in the induced metric) on which $\kappa_g > \pi^2/4r^2$ then we obtain non-existence of invariant tori for all large enough $[P]$.
This can be quite a strong condition, however, e.g.~to make a big bump out of part of a round sphere, one needs to take a whole hemisphere.  

More relevant is to find points in $T^*\Sigma$ whose orbits have conjugate points and thereby restrict the possible invariant tori.  A powerful way to do this is to construct a conefield such that the tangent plane to any invariant torus has to lie in the cones, as presented in \cite{M}.  Points where the conefield is empty correspond to those whose orbits have conjugate points.  The advantage of the conefield formulation is that it permits elimination of further points, namely those from which all tori satisfying the conefield bounds are forced to enter a non-existence region.  This was called the ``killends extension'' of the conefield method.  The reader is referred to \cite{M} or the recent application to magnetic fields in \cite{MM} for details.

\section{Elimination in rotational transform}

The above method finds regions in the phase space $T^*\Sigma$ through which the pressure-jump Hamiltonian has no invariant tori.  It may be desired, however, to present the results in the parameter space of pressure jumps $[P]$ and rotational transform $\iota$.  It is easy to read off $[P]$ as the value of the Hamiltonian (which we take greater than $-V_{\min}$ as before).  To decide which values of $\iota$ are eliminated for given $[P]$ requires further work.  Adapting \cite{M87} and the subsequent \cite{J}\footnote{Note that the comparison at the end of \cite{J} is unfair:~the result of \cite{MP} was that the standard map has no invariant circles of any rotation number for all $k \ge \frac{63}{64}$, not just the single value $k= \frac{63}{64}$, and the method of \cite{MP} could have been applied to a single parameter value much closer to $k_c$ if desired.  Bounding orbits rigorously for a single parameter value can be done much more precisely than for an interval of parameter values, and \cite{J} acknowledged that he had not achieved the latter with sufficient accuracy for interesting results.} to continuous time, here is a way to achieve it, though it can doubtless be streamlined.

The idea is to look for orbits that get ``out of order''.  The projection to $\Sigma$ of an orbit on an invariant torus never crosses itself.  In contrast, the projection of an orbit in the chaotic sea around islands typically spends some time with a higher rotational transform and then some time with a lower one and so it does cross itself.  A consequence of these self-crossings is that there can be no invariant torus with rotational transform between the maximum and minimum of the orbit's finite-time rotational transform, suitably defined.

To formalise this, the projection of an orbit has a set $S$ of ``homology directions'', which is the analogue of rotation interval for discrete time.  
The first homology group $H_1(M,\R)$ for a manifold $M$ is the set of equivalence classes of oriented closed curves in $M$ modulo boundaries of two-dimensional subsets.  For the case of the torus $\T^2 = \R^2/\Z^2$, $H_1(\T^2,\R)$ is a 2D vector space over $\R$ where the basis can be taken to be any pair of classes of closed curves making translation vectors in $\Z^2$ with determinant $\pm 1$.  Unless otherwise specified, we shall take basis vectors $(1,0)$ and $(0,1)$.
Following \cite{Fr}, a {\em homology direction} for a manifold $M$ is a point in $D_M = H_1(M,\R)/\R^+$, under the equivalence relation $h \sim k$ if there is $r>0$ such that $k=rh$.  $D_M$ is topologically the disjoint union of a sphere $\S^{n-1}$ (with $n$ the dimension of $H_1(M,\R)$) and a point $0$.  For the torus, $n=2$.  Given an orbit in $T^*\Sigma$ we define its {\em set of homology directions} $S$ to be the set of limit points in $\S^1 \cup \{0\}$ of the homology directions of closed curves formed by closing segments of the projection of the orbit to $\Sigma$ of length not going to 0 by paths of length going to 0.\footnote{One could define homology directions for orbits in the energy level of $T^*\Sigma$ (where there is an extra dimension of homology from the circle of possible tangent vectors at a point) instead of their projections to $\Sigma$ but we don't consider it useful in the present context.}
For an orbit on an invariant torus, $S$ is a point $(\cos\theta,\sin\theta)$ with $\tan\theta$ being the rotational transform and the ambiguity between $\theta$ and $\theta+\pi$ is resolved by the direction of increasing time.  For any contractible closed geodesic, like around the neck of a big bump, $S = \{0\}$. 

If there is a contractible closed geodesic $\gamma$ then we immediately deduce that there are no invariant tori, because an invariant torus would have an orbit whose projection comes tangent to $\gamma$, but from a given point and velocity vector there is a unique geodesic, giving a contradiction (much more can be deduced about the dynamics, see \cite{DM}).  I expect the argument can be extended to rule out all invariant tori if $0$ is in the set of homology directions for some orbit.  Assuming this, let us restrict attention from now on to the case where for all orbits, $0 \not\in S$.  In that case, $S$ is a single point or a closed arc on $\S^1$.  

If there is an orbit whose set of homology directions is an arc then there is no invariant torus with homology direction in the interior of that arc.  This is because in the 3D energy level in $T^*\Sigma$, an invariant torus of rotational transform $\iota$ implies one with the same rotational transform but the opposite homology direction, by evenness of $H$ in $p$; then any orbit who set of homology directions does not contain $0$ has its set of homology directions constrained to one of the two semicircles between the homology directions for the two invariant tori. 
Thus by building up a set of excluded arcs, one can find restrictions on the possible rotational transforms of invariant tori.

To find an excluded arc, say an orbit has a positive crossing if its projection to $\Sigma$ crosses itself after some positive time at an anticlockwise angle.  This gives a segment of the projected orbit that forms a closed curve, so advancing by some non-zero integer vector $(m_+,n_+)$ on the universal cover.  Suppose it also has a negative crossing, meaning it crosses itself with a clockwise angle after some positive time, and the segment makes a closed curve advancing by some non-zero $(m_-,n_-)$.  If $n_+m_- > m_+n_-$ then we deduce there are no invariant tori with the same $[P]$ with homology direction in the arc between those two rational directions.  By examining longer orbit segments one may be able to produce a sequence of wider arcs.  

This is a continuous-time version of the method of \cite{M87} for discrete time.\footnote{Note that one reader complained the algorithm in the appendix didn't work, so perhaps there are typographical errors that need correcting.}  The idea was adapted by \cite{J}.
Analogously to those references, the method works because the pressure-jump Hamiltonian is positive-definite in the momenta.
The reason for the slightly convoluted formulation in this case is that the set of possible homology directions is a circle plus a point, in contrast to the set of possible rotation numbers for twist maps which is $\R$ (or an interval), so there is only a cyclic order on the circle, not a total order.

One could skip the step of looking for conjugate points and just apply the idea of \cite{M87, J} directly.  I would argue, however, that it is more efficient to find conjugate points than orbits with non-trivial arc of homology directions, and so it is best to use the conjugate point method first to identify suitable orbits for computation of arc of homology directions.  Note that although orbits in islands get out of order they are not useful for getting non-trivial set of homology directions.  The useful orbits are those in the chaotic layers around islands.

It would be great to test this extension, to deduce intervals of rotational transform with no solutions.

An alternative could be to present the results in the space of periods for the magnetic potential.  It would be interesting to develop that approach too.

\section{Integrable metrics on tori}
We now address the question whether there are other cases besides axisymmetric ones, of interfaces for which the Hamilton-Jacobi equation has solutions for almost all winding ratios, at least for some pressure jump.
This can be investigated in the context of Riemannian metrics, by the above reduction to the Maupertuis metric for given pressure jump.

Say a metric on $\T^2$ is {\em completely integrable} if the unit cotangent bundle $T_1^*\T^2$ is foliated by invariant tori of the geodesic flow.
The only metrics on tori that are completely integrable are the flat metrics \cite{BI}.
This is impossible for a $C^2$ toroidal interface $\Sigma$ in Euclidean space.  To prove it, enclose $\Sigma$ in the smallest possible sphere, then at any point of contact the Gaussian curvature of $\Sigma$ is at least that of the sphere, hence non-zero, so the induced metric on $\Sigma$ is not flat.\footnote{Nevertheless, there are $C^1$ embeddings of flat tori in Euclidean space \cite{BJLT}.}

One can generalise the concept of integrability, however, to allow foliations with finitely many singular leaves.  Then the tori of revolution of generic closed curves are included.  Let $x$ be an arc-length parametrisation of a simple closed $C^1$ curve $\gamma$ in a poloidal section.  The surface of revolution has metric $$ds^2 = dx^2 + r(x)^2 d\phi^2,$$ 
where $\phi$ is toroidal angle and $r(x)$ is the distance to the rotation axis.  The geodesic flow has integral $p_\phi = r(x)^2 \frac{d\phi}{ds}$.  The derivatives of $p_\phi$ and $H$ are linearly independent except along the circles of critical points of the function $r$, which we suppose to be finite in number.  Thus the geodesic flow has an invariant torus of each winding ratio except 0.  The invariant torus of winding ratio 0 is replaced by finitely many separatrices around integrable islands.  For the example where $\gamma$ is a round circle, including analysis of its conjugate points, see \cite{W}.

The point of this section is to highlight that there is a larger class of integrable metrics on tori, in this weaker sense, known as {\em Liouville metrics}.
They have the form $$ds^2 =  (f_1(x_1)+f_2(x_2))(c_1^2 dx_1^2+c_2^2 dx_2^2)$$ 
with $c_1, c_2, \min f_1 + \min f_2 > 0$. To see the integrability, note that it is the Maupertuis metric for 
$$H = \frac12 \left(\frac{p_1^2}{c_1^2} + \frac{p_2^2}{c_2^2}\right) -f_1(x_1)-f_2(x_2)$$ 
on $H=0$, which is separable and hence integrable.  There are at most two rotational transforms for which invariant tori do not exist, namely $0/1$ and $1/0$.
To see that the Liouville class includes the tori of revolution, change the parametrisation of $\gamma$ to $y$ with $dy = dx/r(x)$.  Then $ds^2 = r(x(y))^2 (dy^2 + d\phi^2)$ which has the Liouville form.

Can non-axisymmetric Liouville tori be realised as interfaces in Euclidean (3-)space?  Every $C^2$ Riemannian metric on a torus is conformally flat, that is, there exist ``isothermal'' coordinates $(x_1,x_2)$ with period 1 such that $$ds^2 = f(x_1,x_2) (A dx_1^2 + 2B dx_1 dx_2 + C dx_2^2)$$ for some positive $\Z^2-$periodic $C^2$ function $f$ and positive-definite quadratic form.  This is a case of Poincar\'e's uniformisation theorem.  For a proof of the local result, see \cite{Ch}; for construction using the Ricci flow, see \cite{Ha}; or just solve $\Delta_g w = \kappa_g$ for a function $w$ on the torus, where $\Delta_g$ and $\kappa_g$ are the Laplacian and curvature for the metric $g$ on the torus, then $h= e^{2w} g$ is the desired flat metric (up to scaling) and $f = e^{-2w}$.   So, given a toroidal interface, the Maupertuis metric for pressure-jump $[P]$ is $$ds^2 = \frac12 ([P]+V(x)) f(x) (A dx_1^2+2B dx_1 dx_2 + C dx_2^2)$$ in these coordinates.  To be a Liouville metric we must be able to make a linear change of coordinates respecting the periodicity to reduce the quadratic form to a diagonal one and then we require the resulting $([P]+V)f$ to be a sum of functions of $x_1$ and $x_2$ separately.  Is there a restriction on embedding such a torus in Euclidean space?

The problem of embedding a torus with a general $C^2$ metric in Euclidean space appears to be open.  One necessary condition is that the integral of the curvature over the subset where it is positive must be at least $4\pi$. 
For a review and a result that can be translated to Liouville metrics with $f_2=0$, see \cite{HL}.


It is conjectured that the Liouville metrics are the only integrable metrics on tori (modulo possible restriction to integrals that are polynomial in the momenta; see \cite{Ko} for early work and \cite{He} for a nice survey), though those flat metrics having no rectangular period-parallelogram should be included too.  If so, all the others lack invariant tori of some set of positive measure of rotational transform.



\section{Conclusion}

The paper has described a method to compute regions in $T^*\Sigma$ through which pass no solutions of the Hamilton-Jacobi equation for the pressure-jump Hamiltonian.  It has also extended the method to compute regions in the parameter space of pressure-jump $[P]$ and rotational transform $\iota$ for which there are no solutions.  With suitable computational work, but expected to be much less than for KAM theory, the region remaining undecided should converge to the set for which there is a solution, thus making the method a constructive tool.
In the other direction, the paper has highlighted a class of integrable metrics on tori that might provide interfaces for which solutions exist for all but two rotational transforms.

\section*{Acknowledgements}
I am grateful to Stuart Hudson for comments.
This work was supported by a grant from the Simons Foundation (601970, RSM).

\end{document}